\documentclass[12pt,preprint]{elsarticle}
\usepackage{amsmath,amssymb}
\usepackage{xspace}
\usepackage{psfrag,graphicx}
\usepackage[labelsep=period]{caption}
\usepackage{booktabs}
\usepackage[dvipsnames]{xcolor}
\usepackage{lineno}
\PassOptionsToPackage{hyphens}{url}
\usepackage[hyphens]{url}
\usepackage{url}
\usepackage[breaklinks,hidelinks]{hyperref}
\usepackage{breakurl}
\hypersetup{colorlinks = false,linkcolor = blue,anchorcolor =red,
            citecolor = blue,filecolor = red,urlcolor = red,
            pdfauthor=author}
     

\journal{Astroparticle Physics}

\begin{document}
\begin{frontmatter}

\title{Performance of the ISS-CREAM calorimeter in a calibration beam test}
\author[1]{H.G. Zhang}

\author[1,4]{D. Angelaszek} 
\author[1]{M. Copley}       
\author[1]{J.H. Han }
\author[1]{H.G. Huh}
\author[5,7]{Y.S. Hwang}
\author[5,10]{H.J. Hyun}    
\author[5]{H.B. Jeon}    
\author[1]{K.C. Kim}
\author[1]{M.H. Kim}
\author[5]{H.J. Kim}
\author[1]{K. Kwashnak}     
\author[1,6]{M.H. Lee\corref{cor1}}
\cortext[cor1]{Corresponding author: mhlee@ibs.re.kr}
\author[1]{J.P. Lundquist}
\author[1]{L. Lutz}         
\author[1]{A. Malinin}      
\author[5]{H. Park}
\author[5,8]{J.M. Park}
\author[1]{N. Picot-Clemente}
\author[1,4]{E.S. Seo \corref{cor2}}
\cortext[cor2]{Corresponding author: seo@umd.edu}
\author[1]{J. Smith}
\author[1]{J. Wu}
\author[1]{Z.Y. Yin}    
\author[1,9]{Y.S. Yoon\corref{cor3}}
\cortext[cor3]{Corresponding author: ysy@kriss.re.kr}

\address[1]{Institute for Physical Science and Technology, University of Maryland, College Park, MD 20742, USA}
\address[4]{Department of Physics, University of Maryland, College Park, MD 20742, USA}
\address[5]{Department of Physics, Kyungpook National University, Daegu 41566, Korea}

\address[6]{Center for Underground Physics, Institute for Basic Science (IBS), Daejeon 34126, Korea}
\address[7]{Korea Atomic Energy Research Institute, Gyeongju 38180, Korea}
\address[8]{Advanced Radiation Technology Institute, Korea Atomic Energy Research Institute, Jeongeup 56212, Korea}
\address[9]{Korea Research Institute of Standards and Science, Daejeon 34113, Korea}
\address[10]{Pohang Accelerator Laboratory, Pohang 37673, Korea}

\begin{abstract}
The Cosmic Ray Energetics And Mass experiment for the International Space Station (ISS-CREAM) was installed on the ISS to measure high-energy cosmic-ray elemental spectra for the charge range $\rm Z=1$ to 26. The ISS-CREAM instrument includes a tungsten scintillating-fiber calorimeter preceded by a carbon target for energy measurements. The carbon target induces hadronic interactions, and showers of secondary particles develop in the calorimeter. The energy deposition in the calorimeter is proportional to the particle energy.

As a predecessor to ISS-CREAM, the balloon-borne CREAM instrument was successfully flown seven times over Antarctica for a cumulative exposure of 191 days. The CREAM calorimeter demonstrated its capability to measure energies of cosmic-ray particles, and the ISS-CREAM calorimeter is expected to have a similar performance. Before the launch, an engineering-unit calorimeter was shipped to CERN for calibration and performance tests. This beam test included position, energy, and angle scans of electron and pion beams together with a high-voltage scan for calibration and characterization. Additionally, an attenuation effect in the scintillating fibers was studied. In this paper, beam test results, including corrections for the attenuation effect, are presented.
\end{abstract}

\begin{keyword}
ISS-CREAM, cosmic rays, calorimeter, calibration, energy response, attenuation
\end{keyword}

\end{frontmatter}

\section{Introduction}
The balloon-borne CREAM and its successor ISS-CREAM were designed to directly measure the cosmic-ray elemental spectra from protons to iron nuclei, including electrons with ISS-CREAM, for energies ranging from a few hundred GeV to $\sim$ 1 PeV~\cite{seo1,seo2}. CREAM completed seven successful operations circumnavigating Antarctica under a long-duration balloon, and its accumulated exposure was approximately 191 days.

The CREAM instrument proved capable of measuring cosmic-ray elemental spectra with energies over a few hundred TeV. Measured elemental energy spectra and relative abundances of primary cosmic ray nuclei were reported~\cite{seo3,seo4,seo5,seo6}.

ISS-CREAM was installed on the International Space Station (ISS) as a small total exposure limited the balloon-borne CREAM experiment's energy spectrum measurements. This installation allowed a greater run time for higher statistics and a higher cosmic-ray energy maximum.

In August 2017, the ISS-CREAM payload was launched and deployed on the ISS, and in February 2019, finished its mission with 546 days of data to be analyzed (more details are in~\cite{seo7,seo8,seo9}). Detailed descriptions of each of the ISS-CREAM sub-detectors are available elsewhere~\cite{seo10,seo11,seo12,seo13,seo14}.

Before the launch, an ISS-CREAM engineering-unit~\cite{seo15}, a copy of the flight unit, was taken to the European Organization for Nuclear Research (CERN) and exposed to electron and pion beams for calorimeter calibration and performance tests. In this paper, we report the analysis results of the calibration tests and summarize the calorimeter performance.

\section{Calorimeter Description}

\begin{figure}[ht]
\centering
\includegraphics[width=100mm]{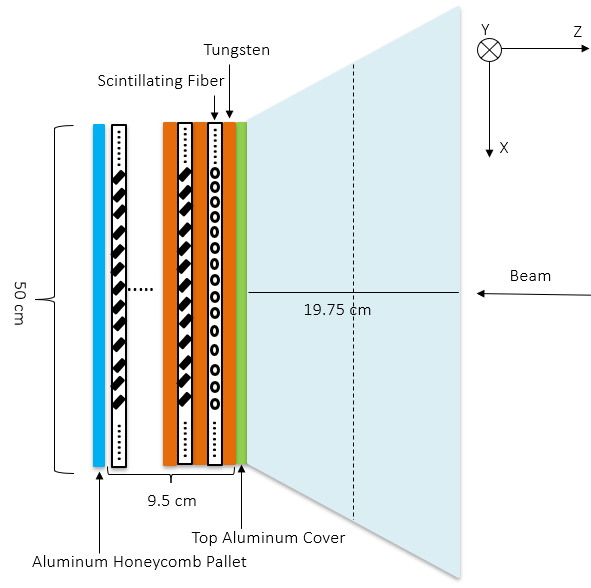}
\caption{\label{cal} A cross-sectional view of the calorimeter and the carbon target in the beam test configuration. The calorimeter comprises 20-layers of tungsten plates and 1000 scintillating fiber ribbons between the layers (not to scale).}
\end{figure}

The ISS-CREAM calorimeter unit consists of the carbon target and the calorimeter made of twenty layers of tungsten plates and scintillating-fiber ribbons.
It is consistent with the CREAM calorimeter in basic design. Figure~\ref{cal} shows a cross-sectional view of the ISS-CREAM calorimeter and the carbon target in the [X, Z] plane. The aluminum cover and calorimeter layers were stacked on top of an aluminum honeycomb pallet.

For compliance with space-launch requirements, the calorimeter was bonded with epoxy to accommodate space transportation's mechanical stresses. More details are available in the ISS-CREAM instrument references~\cite{seo10,seo11,seo12,seo13,seo14}.

The carbon target initiates hadronic showers upstream of the calorimeter with two densified graphite blocks with a density of $\rm 1.92~g/cm^3$ and a total thickness of 19.75~cm, which corresponds to 0.5 nuclear interaction lengths and one radiation length. A significant portion of the electromagnetic shower core and the shower maximum in the longitudinal shower profile is contained in the tungsten/scintillating fiber ribbon layers.

There are twenty 3.5 mm thick $\rm 50~cm \times 50$ cm tungsten plates placed below the target. Each tungsten plate corresponds to one radiation length. A layer with fifty fiber ribbons is placed between each consecutive tungsten layer covering the same area as the tungsten plate. Every other ribbon is mated with light guides on one side of the calorimeter and the other half are mated to the opposite side. The height of the calorimeter with twenty layers of the tungsten plates and scintillating-fiber ribbons is $\rm 9.5~cm$.

Each fiber ribbon is made of nineteen 0.5 mm diameter scintillating fibers with a width of  $\rm 1~cm$. The scintillation light generated in the fibers by charged particles from the shower is transmitted to a hybrid photodiode (HPD) via a light mixer and a bundle of transparent fibers~\cite{seo16}. The light signal is divided into three sub-bundles; low, mid, and high ranges by different numbers of transparent fibers (42, 5, and 1 respectively). Combining the three optical sub-ranges gives enough dynamic range to cover the energies of the measured cosmic-ray particles~\cite{seo16}.

After the conversion of photons to photo-electrons in the HPD's photo-cathode, they are accelerated between the photo-cathode and a pixelated silicon diode to create electron-hole pairs. The pairs are read out by a charge amplifier and sent to both a fast-shaping circuit for a trigger and a slow-shaping circuit to be digitized with an analog to digital converter (ADC) chip (details can be found in~\cite{seo16}).

\section{Calorimeter Calibration}
\label{s-position-scan}

\begin{figure}[ht]
\centering
\includegraphics[width=90mm]{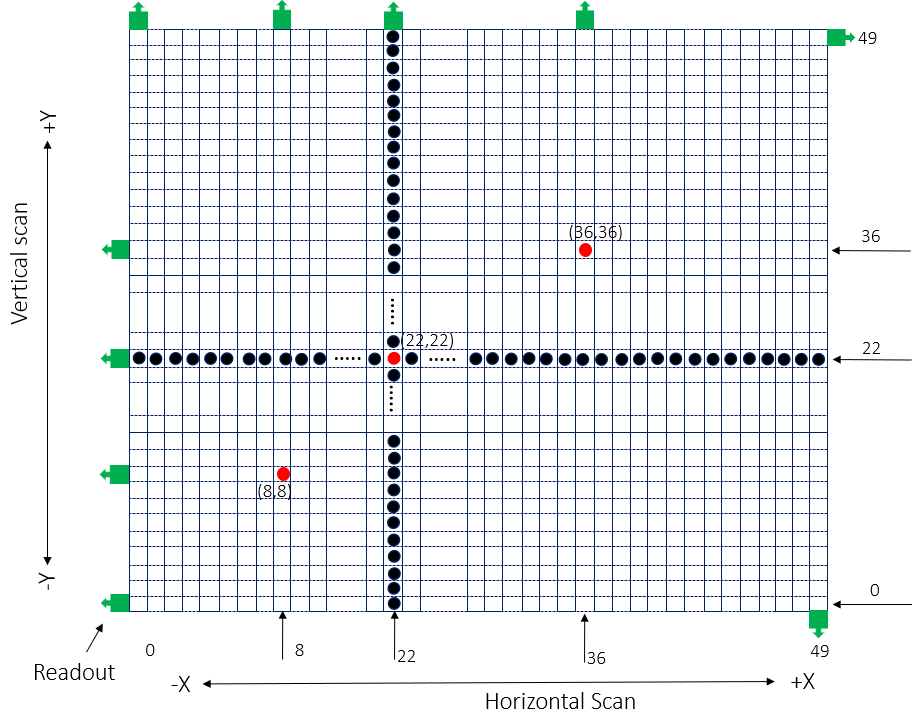}
\caption{\label{scan} Part of the first two-layers of the calorimeter ribbon layout, and the beam hit-position illustration. Layers with ribbons aligned along the $\rm \pm Y$-direction are exposed to beams by moving the hit-position along the $\rm \pm X$-direction. Layers with ribbons aligned along the $\rm \pm X$-direction are exposed to beams by moving the hit-position along the $\rm \pm Y$-direction. The filled black circles represent the incident electron beam spots. The filled red circles represent incident beam spots that are studied in Section~\ref{s-response-resolution}.}
\end{figure}

\begin{figure}[ht]
\centering
\includegraphics[width=90mm]{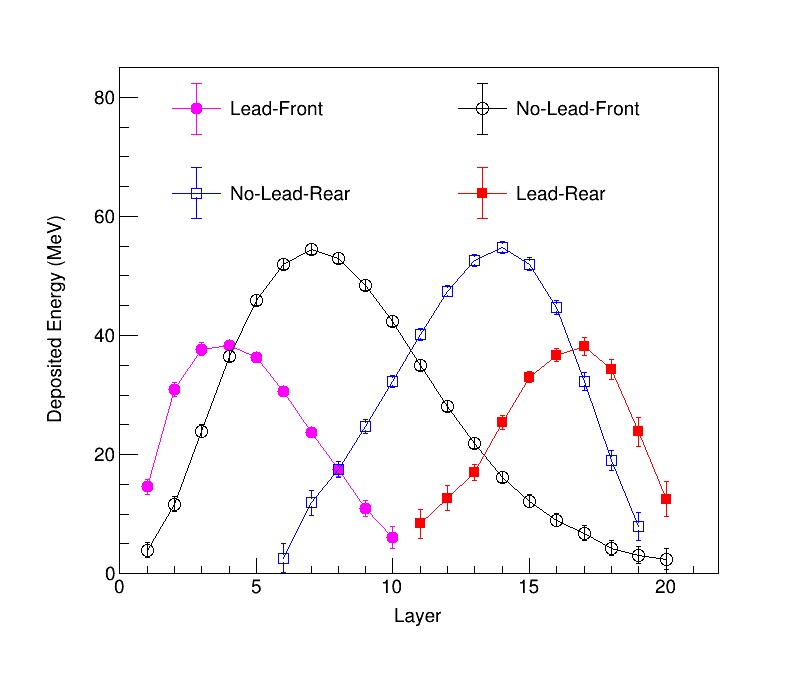}
\caption{\label{averaged} The energy deposit in scintillating fiber ribbons from Monte Carlo (MC) simulation using 150 GeV electron beams for four detector configurations, ``Lead-Front''(filled magenta circles), ``No-Lead-Front''(open black circles), ``No-Lead-Rear''(open blue squares), and ``Lead-Rear''(filled red squares).}
\end{figure}

\begin{figure}[ht]
\centering
\includegraphics[width=90mm]{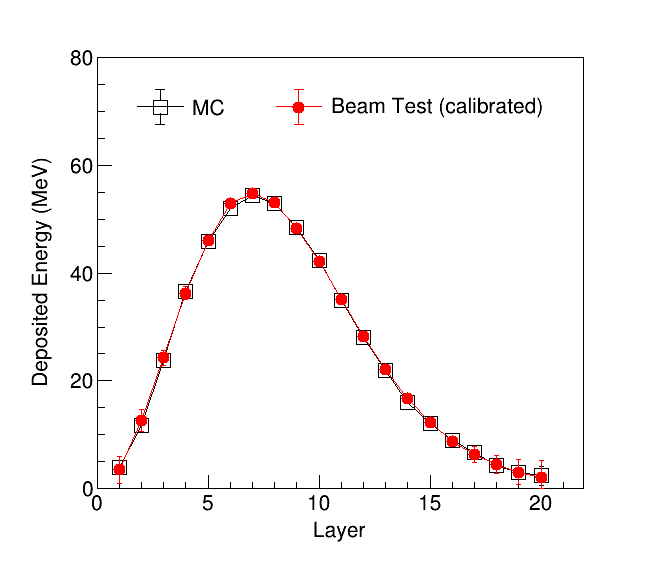}\caption{\label{fit-electron} A comparison of energy deposition in each layer by a 150 GeV electron beam for calibrated data (filled red circles) to that from MC simulation (open black squares).}
\end{figure}

\begin{figure}[ht]
\centering
\includegraphics[width=90mm]{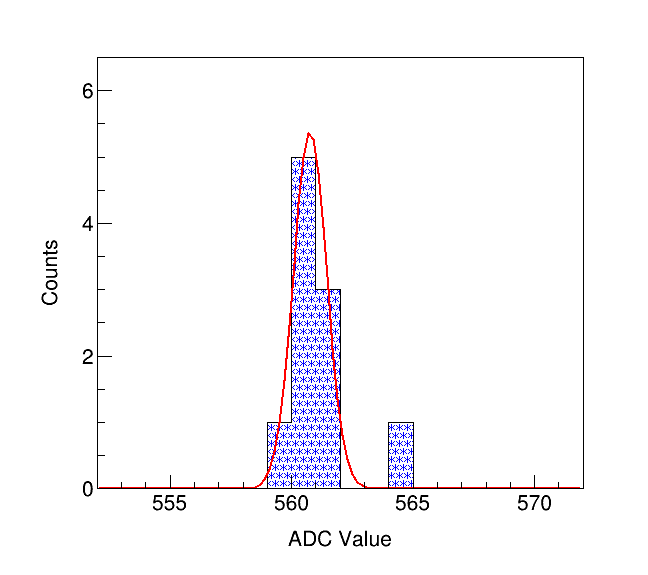}
\caption{\label{adc-stability} The signal response distribution of ribbon 22 in layer 8 for 10-times repeated measurements in both X and Y directions (and a Gaussian fit in red). }
\end{figure}

The purpose of the calorimeter calibration is to obtain the numerical factor required to convert the raw ADC signal values to the measured energy in MeV for each ribbon. An initial calibration factor can be estimated by the ratio of energy deposits in the Monte Carlo simulation to each ribbon's beam response. Additional corrections to the calibration factor were made, including attenuation corrections and gain corrections.

For the initial calibration factor estimation, each ribbon's response was measured with electron beams in the Z direction by shifting the beam position in the X- or Y-direction.

We used a $\rm 150~GeV$ electron beam to calibrate each ribbon. The beam
had a Gaussian distribution profile with a standard deviation of $\sim$4.6~mm \cite{sbt}. As shown in Figure \ref{scan}, ribbons in the 22${\rm nd}$ position from the -X side of odd layers in the calorimeter were exposed to beams. Next, the calorimeter was moved by $\rm 1~cm$ along the X-direction to expose the next ribbon center to the beam. Measurements were repeated for all fifty ribbons in the odd layers. Similarly, the calorimeter was moved by  $\rm 1~cm$ along the Y-direction to expose the even layer ribbon centers to the beams.

Four detector configurations were implemented to measure the ribbon response throughout the calorimeter. The ribbons in the top four layers and the bottom four layers were calibrated using measurements in the ``Lead-Front'' and ``Lead-Rear'' configurations, respectively. The ribbons in layers 5-10 and 11-16 were calibrated with measurements in the ``No-Lead-Front'' and ``No-Lead-Rear'' configurations, respectively.

In the ``No-Lead-Front'' configuration, the electron beam was injected into the front of the graphite target from the +Z-direction.  The ``No-Lead-Rear'' configuration is similar; however, the electron beam was incident on the bottom of the detector from the -Z-direction by rotating the detector. In the ``Lead-Front'' configuration, 2.5~cm-thick lead bricks were placed in front of the graphite target to place the shower maximum in the first few layers. However, the "Lead-Rear" configuration has the lead bricks placed with the instrument rotated to place the shower maximum in the bottom layers. Besides, two tungsten plates were placed between the honeycomb pallet and the lead bricks to compensate for the material above the tungsten-scintillating fiber stack.

Monte Carlo (MC) simulations were performed with the four detector configurations, and the results are shown in Figure~\ref{averaged}. It can be seen that the top and bottom layers have relatively small signals in the ``No-Lead-Front'' and ``No-Lead-Rear'' configurations. To have larger signals at the top and bottom layers, lead bricks were positioned in front of the calorimeter unit.

A pre-selection cut requiring events with sufficient total energy deposit, to remove hadron particle (i.e., pions) contamination, was implemented to select electron events. Figure~\ref{fit-electron} shows perfect agreement between the energy deposit in every layer from the simulation and the beam test (after calibration).

The variation of signal responses at each position was carefully checked in this work and shows good stability. As illustrated in Figure \ref{adc-stability}, the distribution of the responses with ten repeated measurements at the point $\rm [22,22]$ in layer 8, shows a narrow distribution. The ADC distribution has a mean of 561.2 and an RMS of 1.2. This small RMS value shows that the measured ADC values are stable.

\section{Attenuation Effect Measurement}
\label{s-attenuation}

When the beam-hit ribbon position is moved along the X- or Y-direction, the ribbon orthogonal to the moving direction (on the targeted layer) is changed as well. However, ribbons parallel to the moving direction (above or below the targeted layer) do not change with the beam position change. For the ribbons located above or below the targeted layer, data were analyzed at 50 positions with 1 cm intervals from one end to the other. The attenuation effect was studied using ribbons laying parallel to the moving direction in the beam-scan data.

\begin{figure}[hbt!]
\centering
\includegraphics[width=90mm]{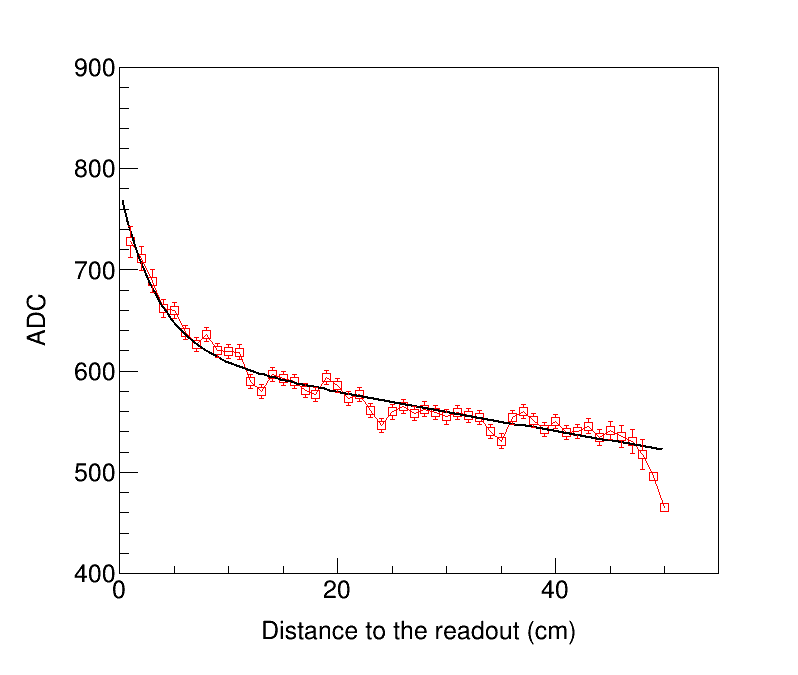}
\caption{\label{layer-8} The signal response of each beam hit position for ribbon 22 in layer 8 when the $\rm 150~GeV$ electron beam moves along that ribbon in the $\rm X$-direction.}
\end{figure}

The transmission of scintillation light in the fiber has two modes. First, if the photon hits a fiber boundary with an incident angle (relative to normal) that is greater than or equal to the critical angle, total reflection happens, and the photon reflects off the surface and remains inside the fiber. Second, if the incident angle is less than the critical angle, whether or not the photon remains inside the fiber depends on the reflection and transmission probabilities. The scintillating fiber's efficiency is determined by the fraction of the photons that arrive at the end of the fiber compared to the total number generated at the hit position. The intensity decrease with distance to the readout $\rm \Delta(x)$ is described by the attenuation function \cite{att1}
\begin{equation}
\rm I= I_{0} \times f(\Delta(x), \epsilon, \lambda_1, \lambda_2)=  I_0\times \epsilon e^{-\Delta(x)/\lambda_1} + I_0\times(1-\epsilon) e^{-\Delta(x)/\lambda_2} ,
\label{fun1}
\end{equation}
where $\rm f(\Delta(x),\epsilon,\lambda_1, \lambda_2)$ is the function related to the transmission and $\rm I_0$ is the intensity at the readout position. The term $\rm I_0\times \epsilon e^{-\Delta(x)/\lambda_1} $ represents transmission with incident angles less than the critical angle, and $\rm \lambda_1$ is the attenuation length for this mode. The term $\rm I_0\times (1-\epsilon) e^{-\Delta(x)/\lambda_2} $ is the transmission behavior of total reflection, and $\rm \lambda_2$ is the attenuation length for this mode. We denote $\rm \epsilon$ as the proportion parameter for the two modes. 
The efficiency at the readout position is normalized to be 1. The attenuation length $\rm \lambda_1$ is shorter than $\rm \lambda_2$, as the light in total reflection travels a longer distance.

\begin{table}
\begin{center}
\begin{tabular}{ |c|c|c|c|c|} 
\hline
Layer & $\rm \epsilon$& $\rm \lambda_{1}(cm)$ & $\rm \lambda_{2}(cm)$ & Efficiency($\%$) \\
\hline
1& $0.10\pm0.01$ & $3.00\pm0.3$ & $247.60\pm12$ & $74\pm3$\\
\hline
2& $0.17\pm0.02$ &$2.05\pm0.4$& $198.44\pm12$& $64\pm3$\\
\hline
3& $0.44\pm0.02$& $1.76\pm0.5$& $282.76\pm14$& $47\pm2$\\
\hline
4& $0.16\pm0.02$ &$3.58\pm0.4$& $174.29\pm10$& $62\pm3$\\
\hline
5& $0.21\pm0.02$& $3.02\pm0.5$& $252.52\pm13$& $64\pm4$\\
\hline
6& $0.23\pm0.02$ &$3.02\pm0.5$& $264.84\pm12$& $63\pm3$\\
\hline
7& $0.25\pm0.03$& $2.94\pm0.5$& $250.06\pm14$& $61\pm3$\\
\hline
8& $0.25\pm0.02$ &$3.02\pm0.4$& $245.12\pm14$& $61\pm3$\\
\hline
9& $0.25\pm0.02$& $2.98\pm0.5$& $252.47\pm12$& $61\pm3$\\
\hline
10& $0.26\pm0.02$& $2.98\pm0.5$& $253.86\pm10$& $60\pm3$\\
\hline
11& $0.22\pm 0.02$& $3.08\pm0.2$& $236.46\pm13$& $63\pm2$\\
\hline
12& $0.22\pm0.03$& $3.03\pm0.4$& $186.90\pm11$& $59\pm2$\\
\hline
13& $0.20\pm0.02$& $3.00\pm0.2$& $251.15\pm12$& $65\pm3$\\
\hline
14& $0.27\pm0.02$& $3.03\pm0.4$& $252.02\pm12$& $59\pm3$\\
\hline
15& $0.30\pm0.03$& $3.02\pm0.3$& $250.44\pm12$& $58\pm2$\\
\hline
16& $0.24\pm0.02$& $3.05\pm0.4$& $242.02\pm12$& $62\pm3$\\
\hline
17& $0.22\pm0.02$& $2.48\pm0.2$& $276.44\pm14$& $65\pm4$\\
\hline
18& $0.31\pm0.02$& $3.04\pm0.3$& $248.28\pm12$& $56\pm2$\\
\hline
19& $0.16\pm0.02$& $1.42\pm0.2$& $222.27\pm13$& $67\pm4$\\
\hline
20& $0.32\pm0.03$& $2.18\pm0.2$& $225.95\pm11$& $54\pm2$\\
\hline
\end{tabular}
\caption{\label{pra-attenuation} 
The best fit parameters ($\epsilon$, $\lambda_1$, $\lambda_2$) and the calculated efficiency at $\rm 50~cm$ from the readout.}
\end{center}
\end{table}

The signal response on each ribbon in the beam-hit position is measured. Figure~\ref{layer-8} shows the signal response distribution of ribbon 22 in layer 8 and is fit by the attenuation function with  $\rm \epsilon = 0.25$,  $\rm \lambda_{1} =3.02~cm$ and $\rm \lambda_{2} = 245.12~ cm$. As we expected, the attenuation length $\lambda_1$ is very short relatively, and the efficiency at the other end of the readout position is about 0.62 (= $ \rm 0.25\times e^{-50/3.02} + (1-0.25)\times e^{-50/245.12}$). Three fit parameters ($\rm \epsilon$, $\rm \lambda_1$, $\rm \lambda_2$) and efficiencies of ribbon 22 in all 20 layers are listed in Table  \ref{pra-attenuation}.  

\begin{figure}[hbt!]
\centering
\includegraphics[width=44.5mm]{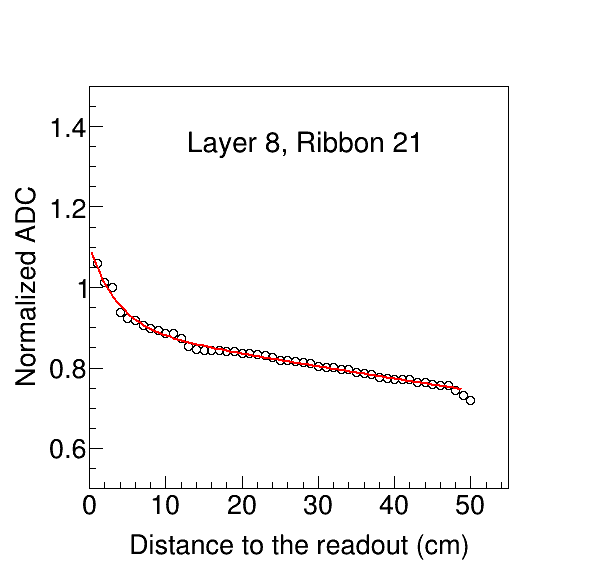}
\includegraphics[width=44.5mm]{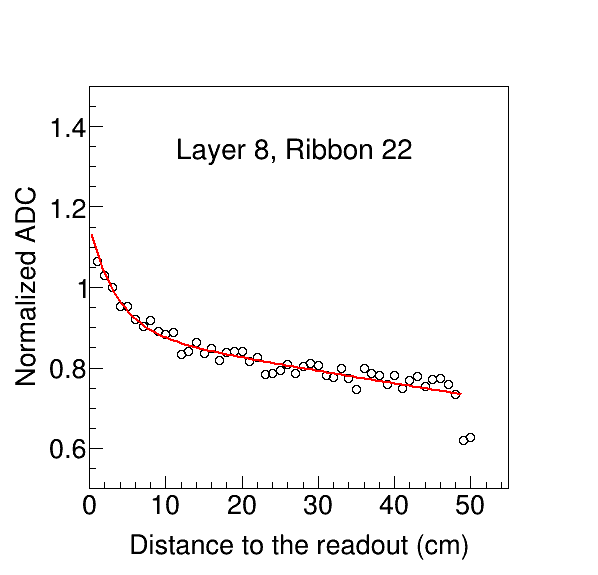}
\includegraphics[width=44.5mm]{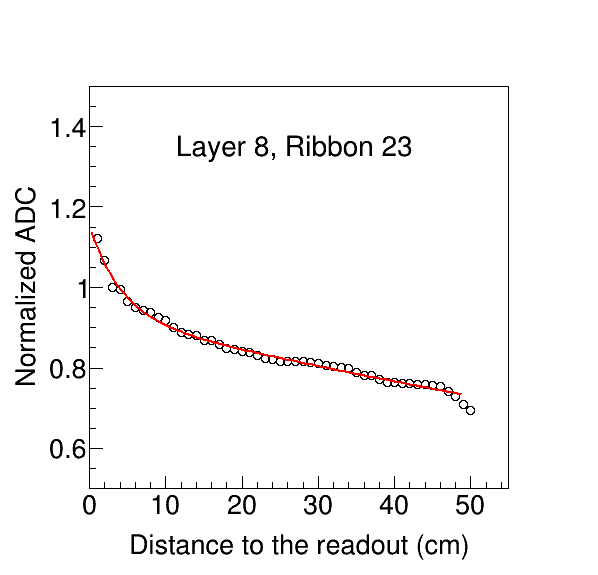}
\caption{\label{att} The signal response distributions of ribbons 21, 22, and 23 in layer 8 are illustrated from left to right. We overlaid the same attenuation function with ($\rm \lambda_1 = 3cm $, $\rm \lambda_2 = 250~cm$, and $\rm \epsilon = 0.25$), showing quite a good agreement. The distributions are normalized to the third points of each ribbon.}
\end{figure}

The most probable attenuation lengths from these fits are found to be $\rm \lambda_1 = 3~cm$ and $\rm \lambda_2 = 250~cm$. Then a new $\rm \epsilon$ is obtained for each layer by fixing the attenuation lengths at the most probable values and leaving $\rm \epsilon$ as a free parameter to apply the attenuation correction for all 1000 ribbons. As an example, Figure \ref{att} illustrates the fits of the signal distributions for ribbons 21, 22, and 23 in layer 8 with $\rm \lambda_1 = 3~cm$ and $\rm \lambda_2 = 250~cm$, and $\rm \epsilon = 0.25$ for all three ribbons.


\section{Attenuation Effect Correction}
\label{s-correction}

The energy response fraction is determined by summing over all calorimeter ribbon energy deposits and dividing by the known beam test incident energy.
With that fraction, the incident energy of a primary particle can be determined through its energy response. However, due to the attenuation effect, the energy deposit is position-dependent,  and this implies the incident energy of the primary particle cannot be properly determined without attenuation correction.

As the attenuation effect is included in the signal measurements for both the beam test and incident cosmic rays, the corrected energy deposit of a primary particle at a random position $\rm [x,y]$ can be expressed as 
\begin{eqnarray}
\rm E_c(x,\Delta(x_c),\Delta(x)) &=& \rm I(x)\times \delta(x_c)\times f(\Delta(x_c))/f(\Delta(x)), \\
\rm E_c(y,\Delta(y_c),\Delta(y)) &=& \rm I(y)\times \delta(y_c)\times f(\Delta(y_c))/f(\Delta(y))
\end{eqnarray}
in even and odd layers, respectively. The signal intensities $\rm I(x)$ and  $\rm I(y)$ come from the measurement of a primary particle at $\rm [x,y]$, and the calibration factors, $\rm  \delta(x_c)$ and $\rm  \delta(y_c)$, come from the beam test measurement at $\rm [x_c,y_c]$. The expressions $\rm f(\Delta(x))$, $\rm  f(\Delta(y))$, $\rm f(\Delta(x_c))$, and $\rm f(\Delta(y_c))$ take the form of Equation (\ref{fun1}), and parameters ($\epsilon$, $\lambda_1$, $\lambda_2$) are set to the values in Table \ref{pra-attenuation}. In fact, $\rm f(\Delta(x_c))$ and  $\rm f(\Delta(y_c))$ represent the correction of the signal responses of the beam test in even and odd layers, and $\rm f(\Delta(x))$ and  $\rm f(\Delta(y))$ represent the correction of the signal response of the primary particle in even and odd layers.

\section{Energy deposit and response with attenuation Correction}

Energy deposits and responses with the attenuation corrections of different positions in the calorimeter are studied using the ``No-Lead-Front'' configuration beam-scan data.

\begin{figure}[hbt!]
\centering
\includegraphics[width=44.5mm]{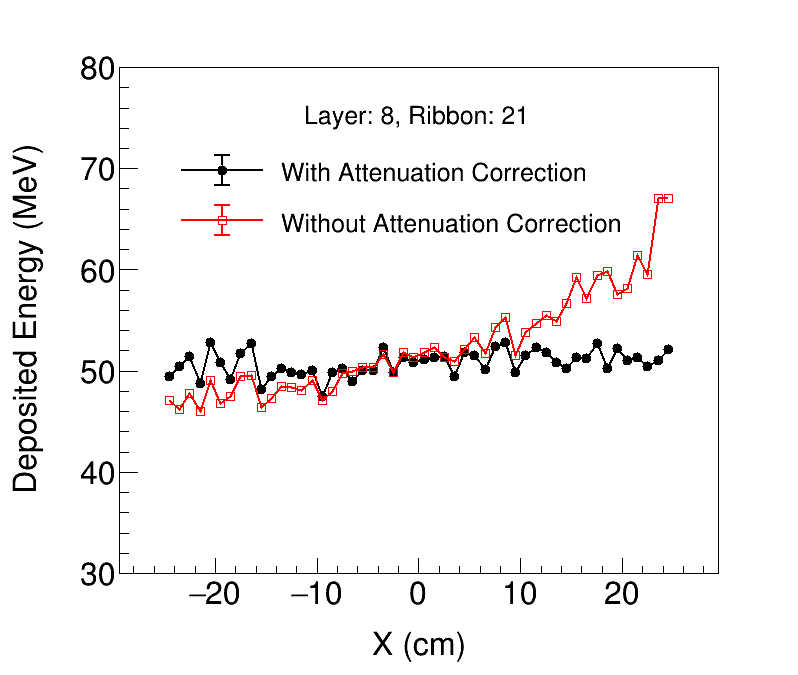}
\includegraphics[width=44.5mm]{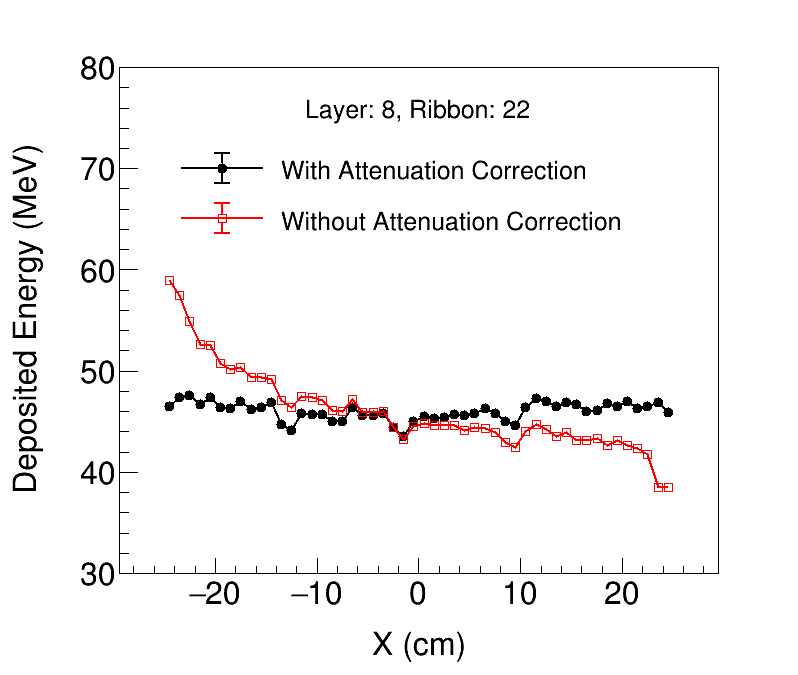}
\includegraphics[width=44.5mm]{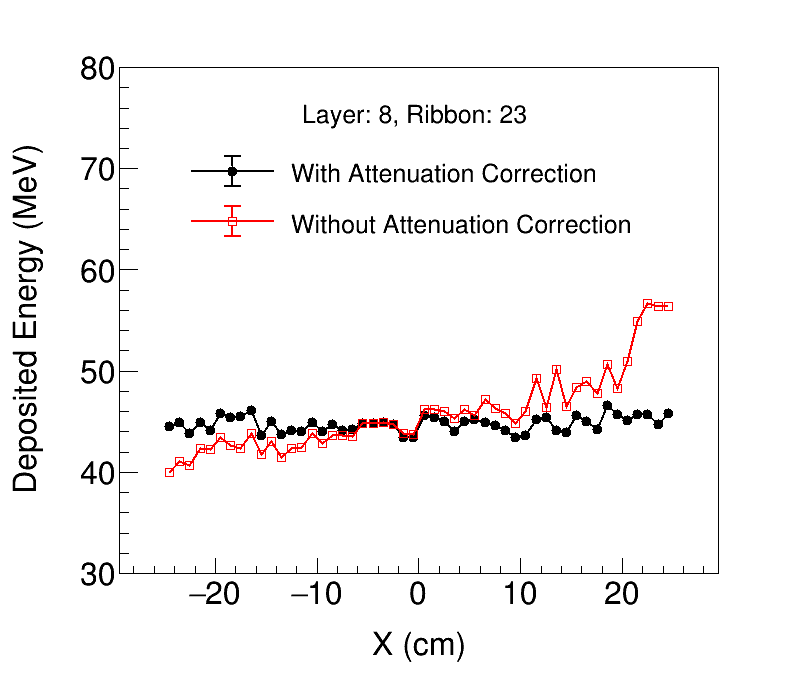}
\caption{\label{layer-8-correction} The energy deposit distribution along three adjacent ribbons in layer 8. We compare the case with attenuation correction (filled black circles) to that without attenuation correction (open red squares). The behaviors of ribbons 21, 22, and 23 are illustrated from left to right.}
\end{figure}

The energy deposit distribution along each ribbon is checked. As an example, we show the energy deposit distributions of ribbons 21, 22, and 23 in layer 8 in Figure \ref{layer-8-correction}. Without attenuation correction, we see the distributions decrease in the X-direction for ribbon 22, and it increases for ribbons 21 and 23. With attenuation correction, the distributions on these three ribbons are position-independent. We find the same results for ribbons in all the 20 layers.

\begin{figure}[hbt!]
\centering
\includegraphics[width=80mm]{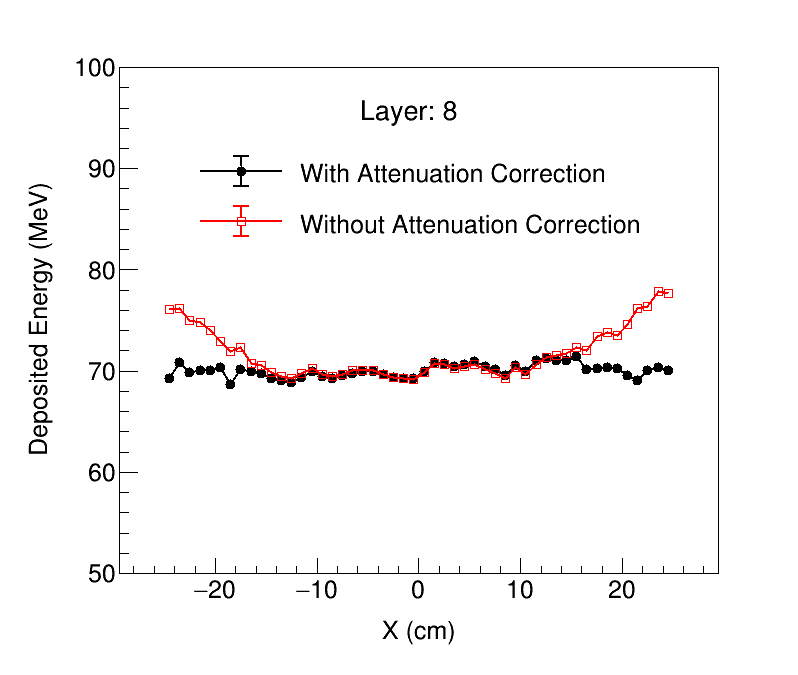}
\caption{\label{layer-8-sum} The total energy deposit in layer 8 for all beam-hit positions. We compare the case with attenuation correction (filled black circles) to that without attenuation correction (open red squares).}
\end{figure}

\begin{figure}[hbt!]
\centering
\includegraphics[width=135mm]{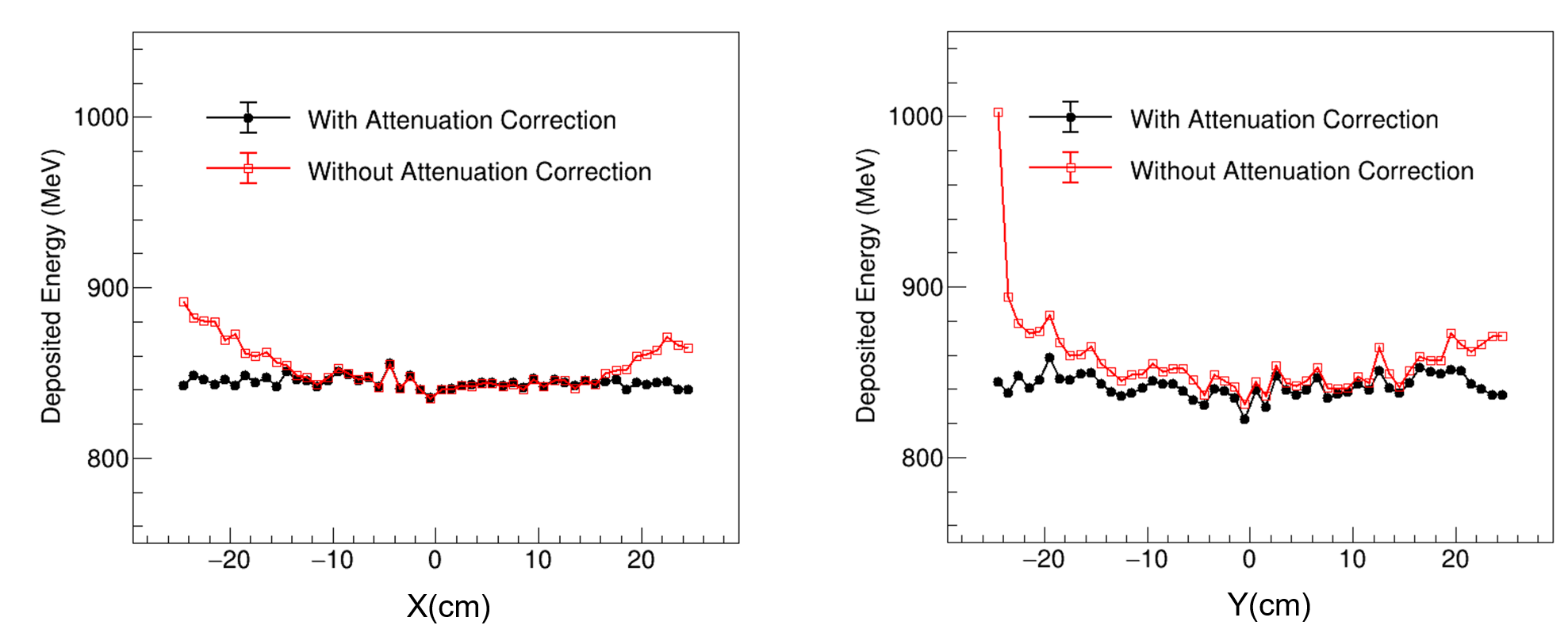}
\caption{\label{response-checking}  The energy response for each beam. We compare the cases with attenuation correction (filled black circles) and without attenuation correction (open red squares) of each beam point. (Left) The beam moves in the X-direction. (Right) The beam moves in the Y-direction. }
\end{figure}

The distribution of the total energy in each layer is checked by summing over the energy deposit in all 50 ribbons for each beam. As an example, we show the total energy distribution of layer 8 in Figure \ref{layer-8-sum}. Without attenuation correction, we find the energies in layer 8 in the middle of the ribbons are lower than those near the edges, while with the attenuation correction, the energies in layer 8 are found to be position-independent as we expected. We find the same result in each layer.

The total energy response, the sum over 1000 ribbons energy deposits in the 20 layers for each beam, is shown in  Figure \ref{response-checking}. We see clearly that the energy responses are position-independent after the attenuation correction. With the same method,  we correct the attenuation effect on for each beam, and study the energy responses for various electron energies, pion energies, electron angles, and electron high voltages in  Section~\ref{s-response-resolution}.

\section{Energy Response and Resolution}
\label{s-response-resolution}

We used scans of electron energy, angle, and high-voltage and a scan of pion energy to characterize the energy response of the calorimeter. The incident beam position is at ribbon 8 in both X and Y directions for the electron energy and pion energy scans, while for the angle and high-voltage scans, the incident position is at ribbon 36 in both X and Y directions.

\subsection{Electron energy response measurements}
\label{s-electron-energy-scan}

\begin{figure}[h!]
\centering
\includegraphics[width=65mm]{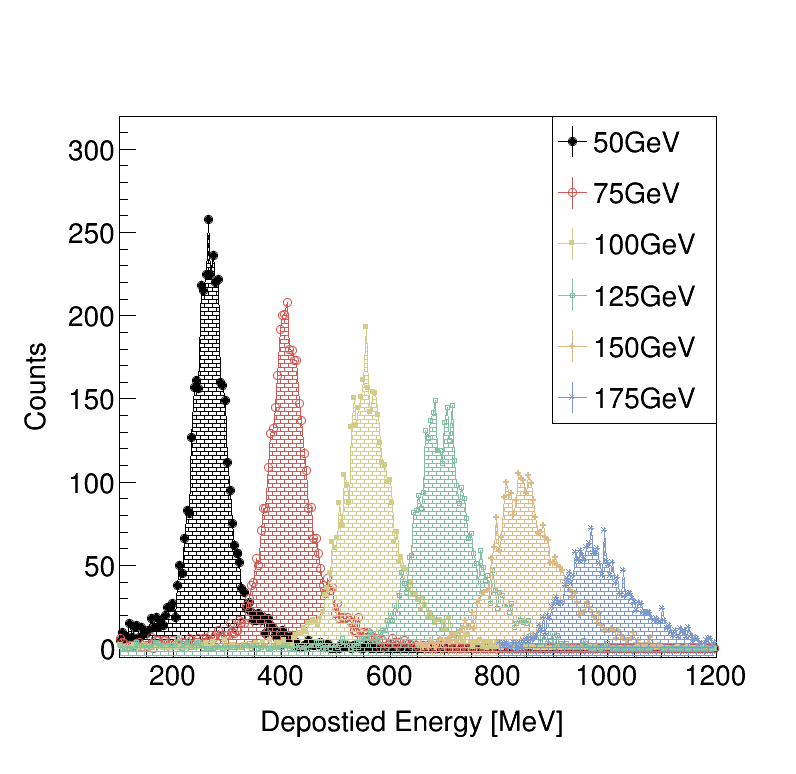}
\includegraphics[width=62.5mm]{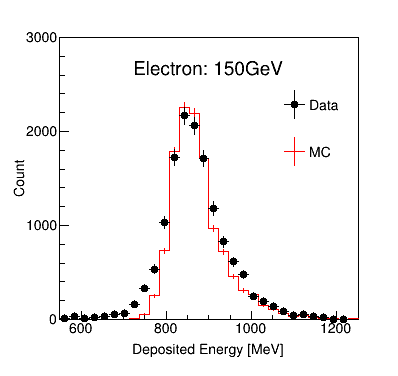}
\caption{\label{energy-distribution} Left: the calorimeter responses to the incident electron beam energies of 50, 75, 100, 125, 150, and 175 GeV, which are distinguished by symbols and colors explained in the inset. 
Right: the energy deposit of beam test data (filled black circles) compared with that from MC simulation (red histogram) for an incident energy of 150 GeV.}
\end{figure}

Figure~\ref{energy-distribution} (left) displays the energy deposits in the calorimeter, for electron beams with increasing energies from $\rm 50~GeV$ to $\rm 175~GeV$, after attenuation correction. Gaussian function fits are used to determine the energy deposits. 


The beam test data are compared with MC simulations taking into account photon statistics and coherent and incoherent electronics noises \cite{smearing}. A good agreement between data and MC can be seen as illustrated in Figure~\ref{energy-distribution} (right) for 150 GeV electrons.

\begin{figure}[hbt!]
\centering
\includegraphics[width=64.5mm]{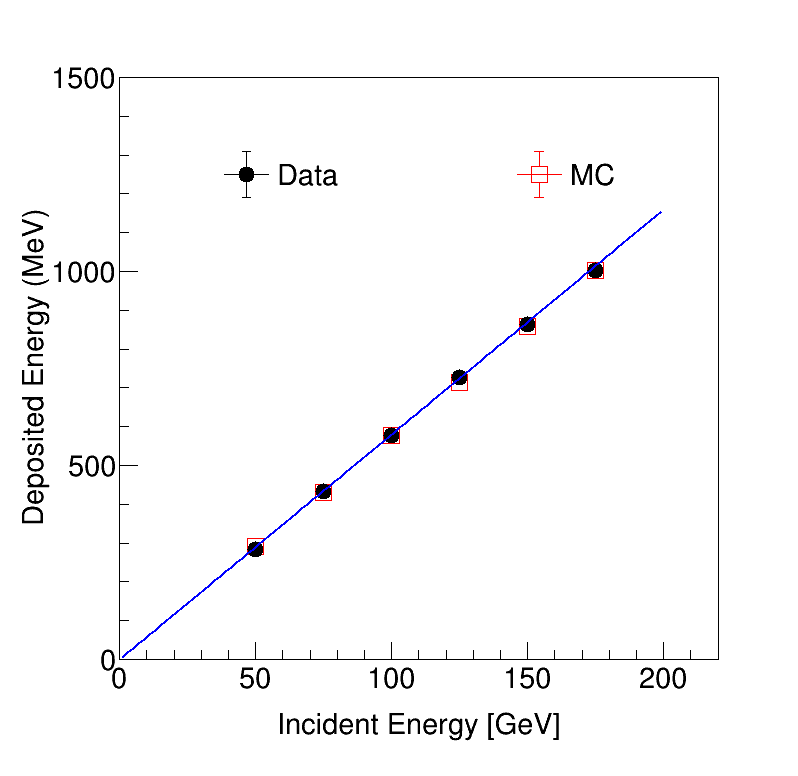}
\includegraphics[width=64.5mm]{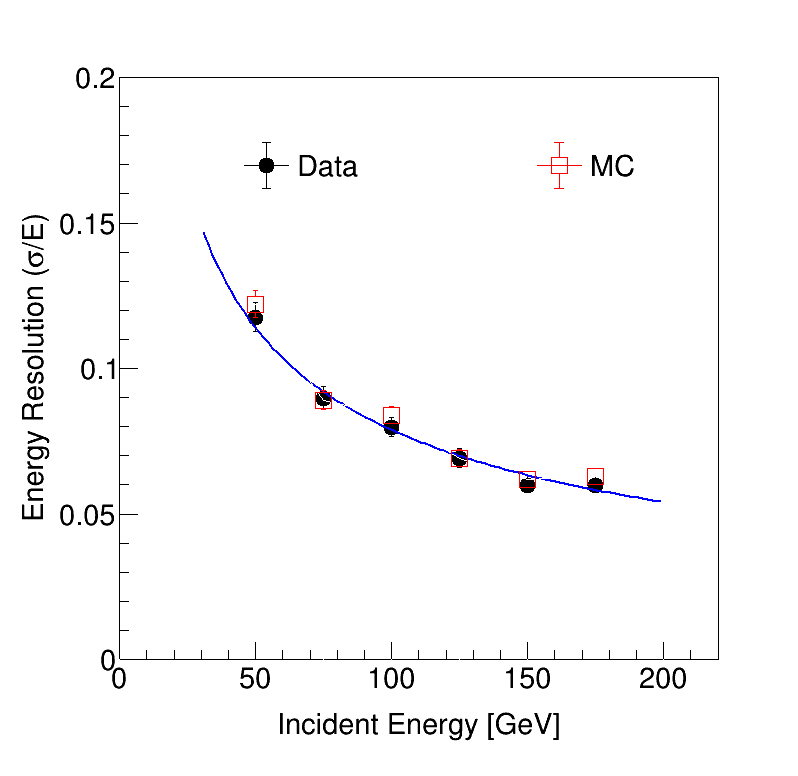}
\caption{\label{energy-resolution} Left: the energy deposits and Right: the energy resolutions as a function of the incident electron beam energies, for both the beam test data (filled black circles) and the MC simulation (open red squares). }
\end{figure}

Figure~\ref{energy-resolution} (left) presents the energy deposit in the calorimeter as a function of the incident electron beam energy. 
The response is linear with increasing electron energy. With attenuation correction, the slope is found to be $\rm 5.74~MeV/GeV$, 
and the total energy deposit with a $\rm 150~GeV$ incident beam is $\rm 860~MeV$. 
They agree with MC simulation results and are consistent with the previous beam test results of the balloon-borne CREAM calorimeters \cite{seo20}.

The energy resolution with respect to the incident electron beam energy is shown in Figure~\ref{energy-resolution} (right), and it is fit to a quadratic sum function of $\rm \sigma_{E}/E(\%)= 85/\sqrt{E(GeV)} \oplus 1.4$. The energy resolution is found to be $\rm 11.4\%$ at 50 GeV and improves with increasing energy, reaching $\rm 6.4\%$ at 150 GeV. The improved energy resolution at higher energy is due to the greater number of shower particles generated in the calorimeter compared with that for low energy electrons. 

\subsection{Pion energy response measurements}
\label{s-pion-energy-scan}

The ribbons' responses were measured with pion beams with energies from $\rm 250~GeV$ to $\rm 350~GeV$ to characterize the calorimeter's energy response to incident hadrons.

To obtain only pion events that interacted in the carbon target upstream of the calorimeter, we performed a pre-selection cut requiring events with significant signals in the first three layers of the calorimeter.

\begin{figure}[hbt!]
\centering
\includegraphics[width=80mm]{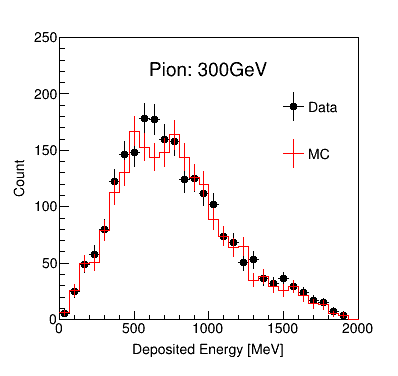}
\caption{\label{pion-distribution} The energy deposit distributions from beam test data (filled black circles) and MC simulation data (red histogram) for 300 GeV pions.}
\end{figure}

The deposited energy distributions from the pion beam test data are also compared with those from MC simulations. As an example, Figure \ref{pion-distribution} shows good agreement between data and MC for $\rm 300~GeV$ pions (with attenuation correction).

\begin{figure}[hbt!]
\centering
\includegraphics[width=64.5mm]{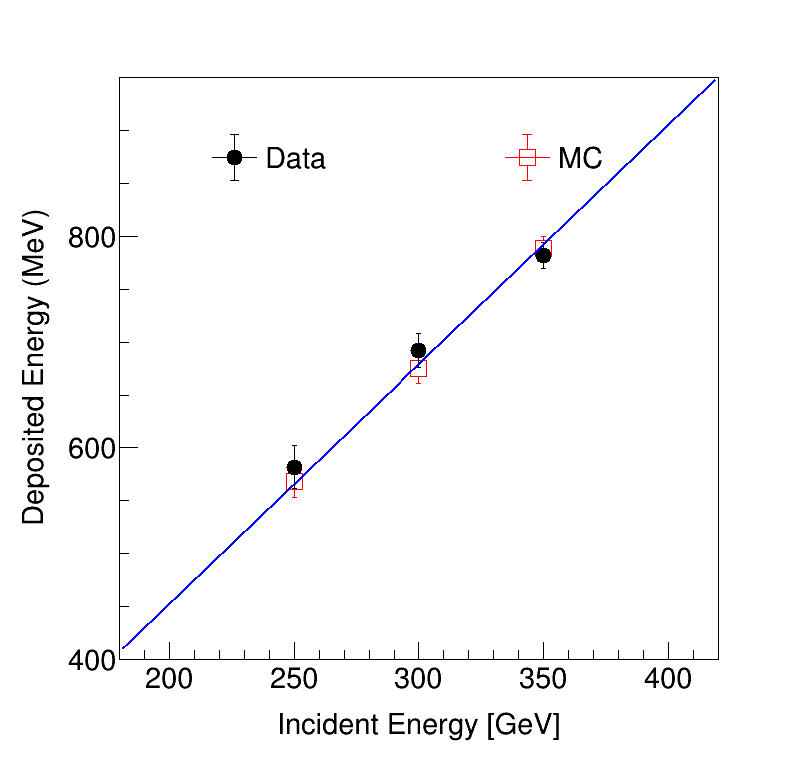}
\includegraphics[width=64.5mm]{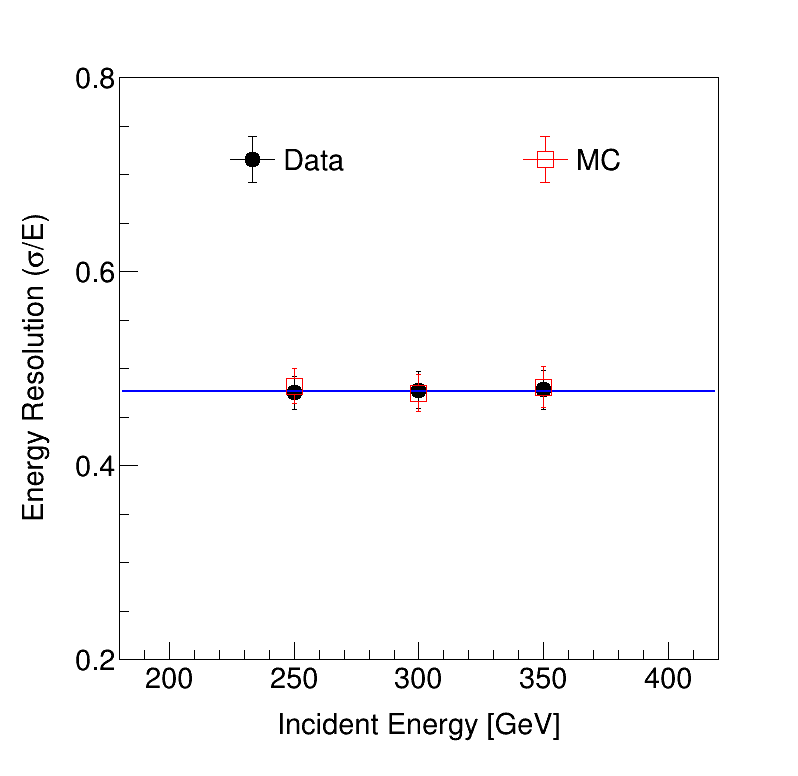}
\caption{\label{pion-resolution} Left: the energy deposits and Right: the energy resolutions as a function of the incident pion beam energies, for both the beam test data (filled black circles) and the MC simulation (open red squares). }
\end{figure}

Figure~\ref{pion-resolution} (left) shows the energy deposit in the calorimeter as a function of the incident pion beam energy. With attenuation correction, the response is found to be linear, with a slope of ~$\rm 1.92~MeV/GeV$, 
which is $\rm 1/3$ the slope of the electron beam. This is consistent with theoretical expectations that when cosmic-ray particles (mainly protons) interact hadronically with the calorimeter, on average approximately $\rm 1/3$ of the energy of the primary nuclei are converted into 
$\rm \pi^{0}$s, which then each rapidly decay into two photons~\cite{seo22}.  

The resolution as a function of the incident pion beam energy is shown in Figure~\ref{pion-resolution} (right). With attenuation correction, the resolution is found to be around $\rm 47.8\%$, which is energy independent.

\begin{figure}[hbt!]
\centering
\includegraphics[width=100mm]{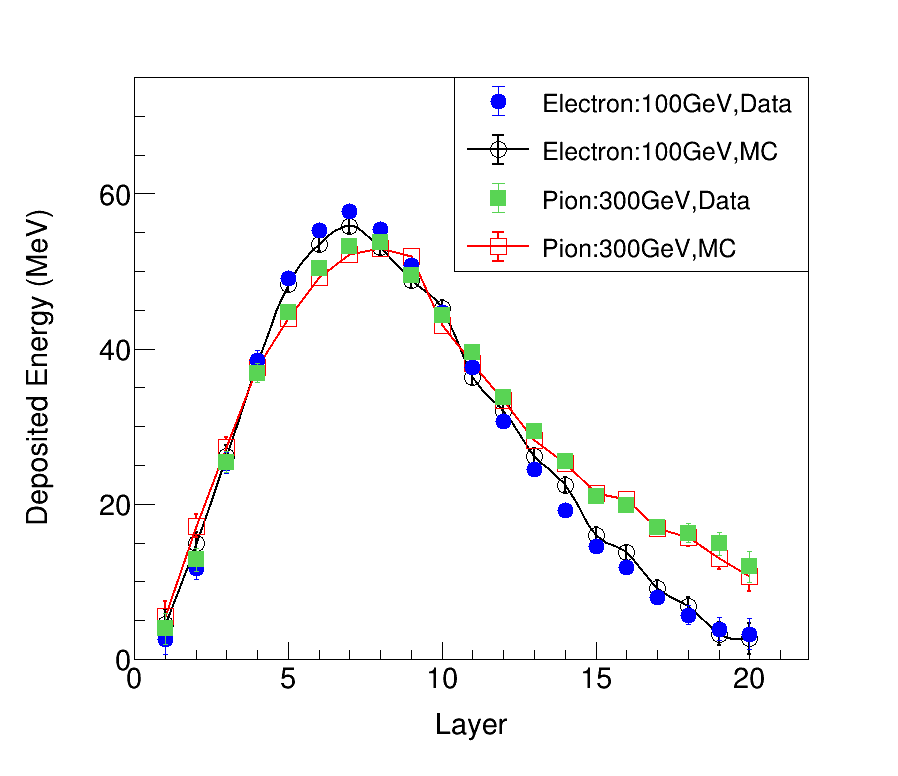}
\caption{\label{mc-data-layer} 
The longitudinal shower development summed over five ribbons in each layer for electron beam test data (filled blue circles) and MC simulation (open black circles), and pion beam test data (filled green squares) and MC simulation (open red squares). The average energy deposit was measured in each layer from 10 data sets repeated in the same positions for both electrons and pions.}
\end{figure}

Figure \ref{mc-data-layer} shows the energy deposit summing over five ribbons in each layer from beam test data and MC simulation for electron beam ($\rm 100~GeV$) and pion beam ($\rm 300~GeV$) respectively. As the pion beam's energy deposit is approximately $\rm 1/3$ of the electron beam, the deposited energies from a $\rm 100~GeV$ electron beam and a $\rm 300~GeV$ pion beam are similar. We show the longitudinal shower profiles of both cases in Fig.~\ref{mc-data-layer} with respect to layer number by summing over the neighboring five ribbons. We see the shower from the pion beam is broader than that from the electron beam. However, for both cases, the primary particles have fully converted their energies into scintillation consistent with the linearity we find in sections \ref{s-electron-energy-scan} and \ref{s-pion-energy-scan}. As shown in Figure \ref{mc-data-layer}, both beam test results agree with those of MC simulations.

\subsection{Electron angle response measurements}
\label{s-electron-angle-scan}

Incident cosmic-ray particles on the detector from various angles in space show a good isotropic approximation. An angle scan was done to characterize the angular response of the calorimeter. Measurements of the electron beam at a fixed energy of $\rm 150~GeV$ was performed with three different incident beam angles of $ \rm 0^\circ$, $\rm 30^\circ$, and $\rm 45^\circ$.

\label{electron-angle-scan}
\begin{figure}[hbt!]
\centering
\includegraphics[width=65mm]{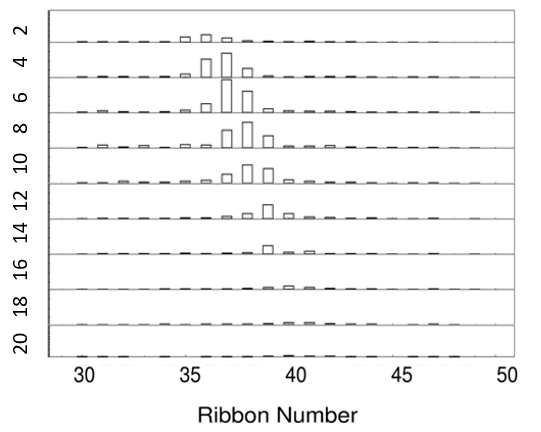}
\caption{\label{angle-track} An event display example of ADC responses of ribbons in even layers using a $\rm 150 ~GeV$ electron beam with an incident angle of $\rm 30^\circ$. The height of the rectangle represents the ADC value of each ribbon. }
\end{figure}

The ADC response of ribbons for the angle scan is shown in Figure~\ref{angle-track}, where the pattern represents the electromagnetic shower from an electron beam with energy of $\rm 150~GeV$ at an incident angle of $\rm 30^\circ$. We see that the electrons traverse the whole detector and produce showers with energy deposited in each layer.

\begin{figure}[hbt!]
\centering
\includegraphics[width=140mm]{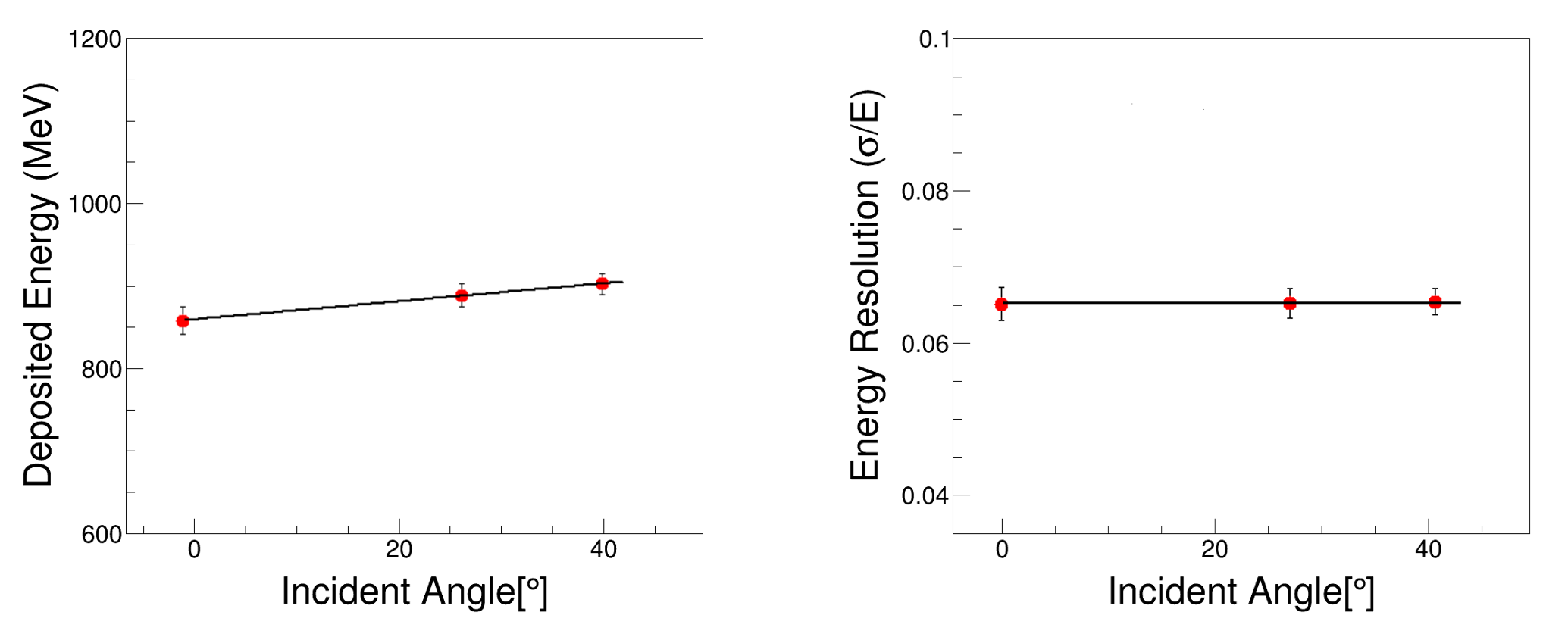}
\caption{\label{angle-linearity} The energy deposit in the calorimeter (left) and the energy resolution (right) as a function of the incident angle using a $\rm 150~GeV$ electron beam. The attenuation correction is applied.}
\end{figure}

Figure~\ref{angle-linearity} (left) shows the energy deposit in the calorimeter as a function of incident angle for a $\rm 150~GeV$ pion beam. With attenuation correction, the response is found to be linear with a fit function of $\rm 1.55\times \alpha + 858$, where $\rm \alpha$ is the angle in degree. There is a $\sim \rm70~MeV$ increase for angles from  $\rm 0^\circ$ to  $\rm 45^\circ$ due to containment inside the calorimeter of more shower particles later in the longitudinal shower development due to the longer available path length. The energy deposit in the calorimeter at $\rm 0^\circ$ is found to be $\rm 858~MeV$ and this result is consistent with what we observed in Section~\ref{s-electron-energy-scan}. The resolutions with increasing incident angles shown in Figure~\ref{angle-linearity} (right) are constant at $\sim \rm 6.5\%$. This result is consistent with what we found in Section~\ref{s-electron-energy-scan}.

\subsection{Measurements with different high-voltage settings}
\label{s-high-scan}

The calorimeter energy response with respect to the applied high-voltage (HV) of the hybrid photodiode (HPD) is studied in this section. This test was necessary for calibration of the balloon-borne CREAM calorimeter, which used different HV values on the HPD during the flight than those used for the beam calibration, to avoid corona discharge damage in the balloon at an altitude of 40 km. For the ISS-CREAM calorimeter, the HPD HV values are the same as those used in the calibration test on the ground.  However, an HV scan test was performed in case any possible situation necessitated different HV values during the flight.

\begin{figure}[hbt!]
\centering
\includegraphics[width=67.5mm]{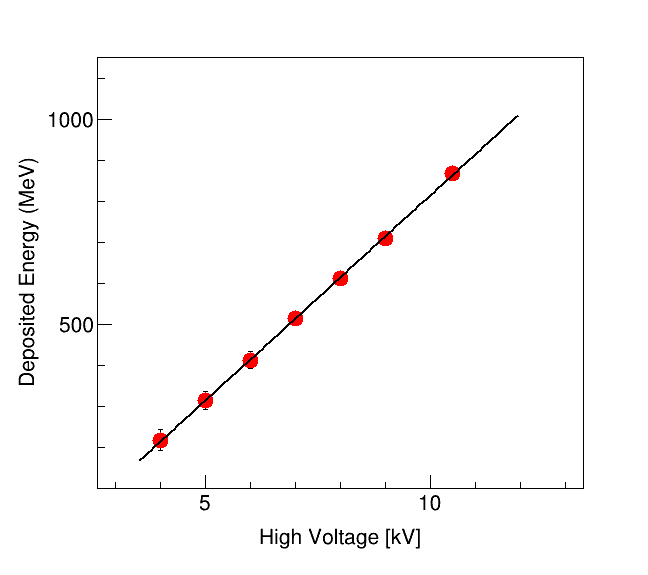}
\includegraphics[width=67.5mm]{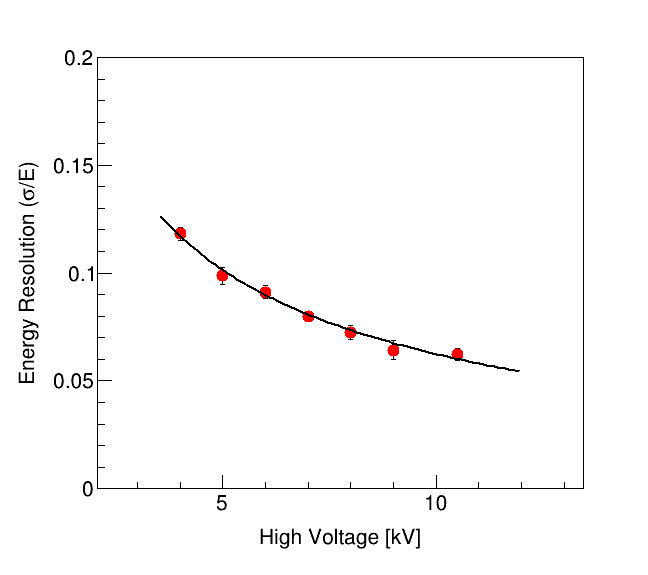}
\caption{\label{hv-resolution} The energy deposit in the calorimeter (left) and the energy resolution of the calorimeter (right) as a function of high voltage values using a $\rm 150~GeV$ electron beam. The attenuation correction is applied. }
\end{figure}

HV gain corrections in the range of 4 kV to 10.5 kV were used to reconstruct flight data's energy deposits. The ADC sum with respect to the HPD's HV was tested at a fixed incident electron beam energy of $\rm 150~GeV$.

Figure~\ref{hv-resolution} (left) shows the total energy deposit as a function of high-voltage. With attenuation correction, the response is found to be a linear function of $\rm 100\times HV - 176$ , where the energy deposit is found to be $\rm 874~MeV$ 
for $\rm 10.5~kV$. This result is consistent with what we found in Section~\ref{s-electron-energy-scan}, where the HV was $\rm 10.5~kV$.

Figure~\ref{hv-resolution} (right) shows the energy resolution of the calorimeter for an electron beam as a function of HV. With attenuation correction, the resolution is fit with a quadratic sum function of $\rm \sigma_{E}/E(H)(\%) =29.8/\sqrt{HV(kV)} ~\oplus ~1$, and the resolution is found to be $\sim \rm 6.2\%$ at $\rm 10.5~kV$. This result is consistent with that found in Section~\ref{s-electron-energy-scan}.

\section{Conclusion}

The tungsten scintillating-fiber calorimeter for the ISS-CREAM was calibrated with electron beams at CERN. The calibration factors of the total 1000 fiber ribbons were estimated by comparing the deposits in the Monte Carlo simulation to the beam response of each ribbon.

In addition to that, we implemented scans of electron and pion energies, angle and high-voltage scans for a 150 GeV electron beam to characterize its energy response. The instrument was exposed to 50,  75, 100, 125, 150 and 175 GeV electrons to characterize the energy response of the calorimeter, and it was also exposed to pions with 250, 300 and 350 GeV. In high-voltage scan, we reconstructed the energy deposits with the HV gain corrections in the range of 4 kV to 10.5 kV, while in angle scan, the measurement of electron beam at a fixed energy of 150 GeV was performed with incident beam angles of $\rm 0^\circ$, $\rm 30^\circ$, and $\rm 45^\circ$.  For all the scans, the attenuation effect due to the hit positions on the fiber ribbons was corrected.

It was confirmed that the energy responses are linear for electron beams with an energy range from 50 GeV to 175 GeV and pion beams with an energy range from 250 GeV to 350 GeV. These were available beams for the detector calibration at CERN and the linearity of measurement energy range was validated with a heavy-ion beam test (see \cite{seo23,seo24}) and MC simulation results up to 1000 TeV (see \cite{seo18,seo25,seo19}).

The energy resolution and the ratio of deposited to incident energy in the calorimeter were found to be consistent with that of the previous balloon-borne CREAM calorimeters. For the electron beam, the energy response slope with increasing incident beam energy is found to be $\rm 5.74~MeV/GeV$, and the energy resolution approaches $\rm 6.4\%$ at 150 GeV. For the pion beam, the slope of energy deposit in the calorimeter versus incident energy is found to be $\rm 1.92~MeV/GeV$, and the resolution, which is independent of the incident energy, is found to be $\rm 47.8\%$. The results of this paper are being used for flight data analysis.

\section{Acknowledgments}

This research was supported in the U.S. by NASA grant NNX17AB41G, in Korea by National Research Foundation grants (2018R1A2A1A05022685 and 2018R1A6A1A06024970) and Ministry of Science and ICT (MSIT, IITP No. 2020-0-015345), and their predecessor grants. The authors also thank the support of CERN for the excellent beam test facilities and operations. H.G.Z. is grateful for the help of Dr. Zu-Hao Li.

\end{document}